%% file: ijcai25.tex
\documentclass{article}
\usepackage{ijcai25}

\usepackage{times}
\usepackage{soul}
\usepackage{xcolor}
\usepackage{url}
\usepackage{multirow}
\usepackage[hidelinks]{hyperref}
\usepackage[utf8]{inputenc}
\usepackage[small]{caption}
\usepackage{graphicx}
\usepackage{amsmath}
\usepackage{amsthm}
\usepackage{amssymb}
\usepackage{graphicx}
\usepackage{booktabs}
\usepackage{algorithm}
\usepackage{algorithmic}
\usepackage{booktabs}
\usepackage[switch]{lineno}
\usepackage{adjustbox}
\usepackage[capitalize]{cleveref}
\crefname{section}{Sec.}{Secs.}
\Crefname{section}{Section}{Sections}
\Crefname{table}{Table}{Tables}
\crefname{table}{Table}{Tables}
\Crefname{figure}{Figure}{Figures}
\crefname{figure}{Fig.}{Figs.}
\Crefname{equation}{Equation}{Equations}
\crefname{equation}{Eq.}{Eqs.}
\crefname{algocf}{alg.}{algs.}
\Crefname{algocf}{Algorithm}{Algorithms}
\usepackage{xspace}
\makeatletter
\DeclareRobustCommand\onedot{\futurelet\@let@token\@onedot}
\def\@onedot{\ifx\@let@token.\else.\null\fi\xspace}

\makeatother
\title{
LensNet: An End-to-End Learning Framework for Empirical Point Spread Function Modeling and Lensless Imaging Reconstruction
}

\author{Anonymous Submission}

\author{
Jiesong Bai$^{1,2}$\and
Yuhao Yin$^2$\and
Yihang Dong$^1$\and
Xiaofeng Zhang$^3$\and
Chi-Man Pun$^1$\and
Xuhang Chen$^{1,4}$\\
\affiliations
$^1$Faculty of Science and Technology, University of Macau\\
$^2$Shanghai University\\
$^3$School of Electronic Information and Electrical Engineering, Shanghai Jiao Tong University\\
$^4$School of Computer Science and Engineering, Huizhou University\\
\emails
xuhangc@hzu.edu.cn
}

\begin{document}

\maketitle

\begin{abstract}
Lensless imaging stands out as a promising alternative to conventional lens-based systems, particularly in scenarios demanding ultracompact form factors and cost-effective architectures. However, such systems are fundamentally governed by the Point Spread Function (PSF), which dictates how a point source contributes to the final captured signal. Traditional lensless techniques often require explicit calibrations and extensive pre-processing, relying on static or approximate PSF models. These rigid strategies can result in limited adaptability to real-world challenges, including noise, system imperfections, and dynamic scene variations, thus impeding high-fidelity reconstruction. In this paper, we propose LensNet, an end-to-end deep learning framework that integrates spatial-domain and frequency-domain representations in a unified pipeline. Central to our approach is a learnable Coded Mask Simulator (CMS) that enables dynamic, data-driven estimation of the PSF during training, effectively mitigating the shortcomings of fixed or sparsely calibrated kernels. By embedding a Wiener filtering component, LensNet refines global structure and restores fine-scale details, thus alleviating the dependency on multiple handcrafted pre-processing steps. Extensive experiments demonstrate LensNet's robust performance and superior reconstruction quality compared to state-of-the-art methods, particularly in preserving high-frequency details and attenuating noise. The proposed framework establishes a novel convergence between physics-based modeling and data-driven learning, paving the way for more accurate, flexible, and practical lensless imaging solutions for applications ranging from miniature sensors to medical diagnostics. The link of code is \url{https://github.com/baijiesong/Lensnet}.
\end{abstract}

\input{figtex/fig1}
\input{figtex/fig2}

\section{Introduction}
Conventional imaging systems that rely on lenses are inherently constrained by factors such as cost, weight, and physical dimensions, which limit their potential for miniaturization~\cite{chen1,chen2,chen3,chen4,chen5,chen6}. In contrast, lensless imaging avoids the use of lenses altogether and instead uses computational algorithms to reconstruct scenes. This lens-free approach not only lowers production costs but also reduces device thickness, supporting the development of ultra-thin imagers. Existing methods frequently employ coded apertures~\cite{ref9} or phase masks~\cite{ref10,ref11} near image sensors to modulate incoming light. These compact and low-cost solutions are well suited for various applications, including miniature sensors and medical imaging. Without lenses, acquired signals become highly multiplexed and require sophisticated computational techniques~\cite{ref12} for accurate reconstruction. Central to these techniques is the Point Spread Function (PSF), which describes how the system responds to a point source and significantly influences imaging performance.

Despite its pivotal role, traditional model-based methods~\cite{ref13,ref14,ref15} often rely on static or calibrated PSF models, which struggle to accommodate real-world challenges such as system imperfections and dynamic scene variations. In classic lensless imaging, the PSF is typically pre-calibrated or inferred through fixed models, leading to rigid assumptions that cannot adapt to practical scenarios. Factors such as changes in lighting, sensor noise, and minor misalignments cause deviations between the assumed PSF and the true PSF, ultimately degrading render quality. Furthermore, the large support of the PSF in lensless imaging presents two major challenges. First, the expanded linear equations resulting from a broad PSF footprint are computationally demanding to store and invert, leading to significant memory and processing overhead. Second, this global multiplexing complicates reconstructions as local image details are intertwined with the global signal. This issue becomes especially detrimental under low-light or noisy settings, where the recovery of fine-scale features is critical, yet hard to achieve.

Recent advances in deep learning~\cite{ref18,ref17,li2024cross,zhu2024test,liu2023coordfill,yao1,yao2,chen7,chen8,chen9,chen10,chen11,chen12,chen13,chen14} combine traditional imaging physics with data-driven neural networks and have achieved substantial progress in reconstructing scenes from heavily multiplexed signals. For example, FlatNet~\cite{ref8} uses paired training data to improve image quality but encounters difficulties in recovering high-frequency details. Similarly, PhoCoLens~\cite{ref7} employs a two-stage pipeline that shows promising performance, yet may blur high-frequency components during its diffusion procedure.

To address these shortcomings, we propose \emph{LensNet}, an end-to-end deep learning framework that effectively fuses spatial and frequency domain information for lensless image reconstruction. Our framework adopts an encoder–decoder architecture, where the encoder features a novel \emph{Coded Mask Simulator (CMS)} module that we design to capture the optical path and raster encoding properties of the lensless imaging process. By modeling the global and multi-scale characteristics of a learnable PSF, this module adapts to diverse imaging conditions without requiring explicit PSF calibration. Furthermore, we incorporate Wiener filtering~\cite{ref12} into our network to refine reconstructions via frequency-domain deconvolution and PSF correction, thereby enhancing noise suppression and edge fidelity. By integrating these customized modules, LensNet takes advantage of both spatial and frequency domain cues to produce robust, high-resolution reconstructions while maintaining global consistency, as showed in \Cref{fig1}. Extensive experiments confirm the efficacy and reliability of our design. Our main contributions are as follows.
\begin{itemize}
\item We propose a novel end-to-end deep learning framework for lensless imaging that dynamically captures multi-scale features in both spatial and frequency domains, substantially improving the fidelity and accuracy of image reconstructions over conventional methods.
\item We introduce a state-of-the-art Coded Mask Simulator that precisely learns the true behavior of the point spread function (PSF), facilitating superior recovery performance across diverse lensless imaging scenarios.
\item Through comprehensive quantitative and visual evaluations, we demonstrate that our proposed LensNet outperforms existing approaches on standard benchmark datasets.
\end{itemize}

\section{Related Work}
\subsection{Traditional Lensless Imaging}
Traditional lensless imaging methods commonly rely on optical encoding techniques. Coded aperture imaging~\cite{ref9} employs a structured mask or aperture to encode incident light, while phase retrieval~\cite{ref10,ref11} reconstructs both the amplitude and phase of the light field from intensity measurements. Although these approaches have shown success in specific scenarios, they often require complex optical setups and computationally expensive reconstruction algorithms that are prone to noise and challenging imaging conditions.

In addition to encoding-based methods, iterative optimization methodologies such as FISTA~\cite{ref2} and ADMM~\cite{ref1} are widely adopted to address inverse problems in lensless imaging. FISTA accelerates sparse signal recovery, while ADMM efficiently solves convex optimization problems by decomposing them into simpler sub-problems. Moreover, ADMM with total variation regularization~\cite{ref5} helps to preserve edges and reduce noise. Despite the effectiveness of these methods, they require careful parameter tuning and remain computationally expensive, particularly under noisy measurements or complex optical systems.

\subsection{Learning-Based Lensless Imaging}
Recent deep learning approaches~\cite{zheng2024smaformer,liu2024depth,ref13,ref21,liu2023explicit} have capitalized on the representational power of neural networks to model complex forward and reverse imaging processes directly from data. Multiple studies~\cite{ref7,ref16,ref22} have validated the effectiveness of these learning-based methods. For example, TikNet and FlatNet~\cite{ref8} use paired training data to improve image quality but struggle to preserve high-frequency details. Similarly, PhoCoLens~\cite{ref7} employs a two-stage processing framework that achieves favorable results; however, its diffusion process blurs fine structures, requiring iterative refinements. ~\cite{ref29} proposes a lensless imaging reconstruction method based on GAN, which achieves robust reconstruction of images captured by multiple cameras without requiring explicit PSF knowledge by jointly performing image reconstruction and PSF estimation. 

Despite these advances, several challenges remain. Precise recovery of fine-scale image details, accommodating fixed or overly simplistic point spread functions (PSFs), and mitigating noise-induced degradation are significant hurdles. Traditional systems often rely on static PSF models that cannot adapt to variations in the scene or account for sensor imperfections, thus hindering image quality. Preserving high-frequency vision cues, such as sharp edges and detailed textures, remains also difficult. Our proposed imaging framework addresses these limitations by introducing a more flexible PSF modeling mechanism and improving the restoration of high-frequency details.

\input{figtex/fig3}
\input{figtex/fig4}
\input{tabletex/1_comp}
\section{Methodology}
In this section, we present a novel lensless imaging framework that jointly exploits both spatial and frequency domain information for high-fidelity image reconstruction. To address the limitations of existing lensless imaging systems, our proposed framework integrates multiple advanced modules specifically designed to overcome inefficiencies and inaccuracies commonly encountered in PSF-based approaches. The remainder of this section describes the overarching network architecture and its constituent modules in detail.

\subsection{Preliminary}
\label{sec:pre}
In optics, the Point Spread Function (PSF), denoted as $\operatorname{PSF}(x, y)$, characterizes how an imaging system distributes a point source of light. Technically, $\operatorname{PSF}(x, y)$ is a two-dimensional or three-dimensional function representing the output signal when a point source is introduced at the input. Coded masks, widely employed as optical encoders, are constructed with specific patterns (e.g., horizontal lines, vertical lines, grids, or random structures) to modulate incident light. By imposing spatial modulation on the optical signal, these masks effectively encode essential information about the scene. In lensless imaging systems, the particular arrangement of the coded mask plays a pivotal role in the definition of $\operatorname{PSF}(x, y)$, thereby influencing the quality and fidelity of the reconstructed images. Hence, the design of the coded mask significantly affects the imaging performance of the system, determining how light is distributed and shaping critical attributes such as resolution, contrast, and image clarity.

The forward lensless imaging model can be mathematically represented as a linear transformation~\cite{ref23}. For an input image, the sensor in a lensless system acquires a blurred observation:
\begin{equation}
    I_{\text{measurement}}(x, y) = I_{\text{object}}(x, y) \ast \operatorname{PSF}(x, y) + \text{noise}(x, y),
\end{equation}
where $I_{\text{object}}(x, y)$ is the original image, $\operatorname{PSF}(x, y)$ denotes the system's point spread function, $\text{noise}(x, y)$ is the sensor noise, and $I_{\text{measurement}}(x, y)$ is the recorded measurement.

By introducing spatial encoding through coded masks, lensless systems modulate the incoming light to create distinctive intensity distributions on the detector. Physically, the coded mask pattern directly affects $\operatorname{PSF}(x, y)$, which emerges in the measurement image via the convolution operation:
\begin{equation}
    I_{\text{measurement}}(x, y) = I_{\text{object}}(x, y) \ast \operatorname{PSF}(x, y).
\end{equation}

Hence, the mask pattern governs the distribution of spatial frequencies, the direction of light propagation, and the overall optical properties, all of which are intrinsically manifested in $I_{\text{measurement}}(x, y)$. Consequently, the choice of coded mask has a profound influence on both the image formation process and the achievable reconstruction quality.

\subsection{Overall Architecture}
The overall architecture of LensNet is illustrated in \cref{fig2}. The framework adopts an encoder-decoder structure, where the encoder processes multi-resolution spatial and frequency domain features in the first stage, and the decoder reconstructs the image in the second stage. We begin by outlining the LensNet framework and then provide a detailed description of its individual components.

Given the measurement image $I_{\text{measurement}}(x, y)$ measured by the sensor in a lensless imaging system, we simplify the analysis by momentarily assuming a noise-free condition. The goal of LensNet is to reconstruct $I_{\text{object}}(x, y)$ from $I_{\text{measurement}}(x, y)$ while satisfying the following equation:
\begin{equation}
    I_{\text{measurement}}(x, y) = I_{\text{object}}(x, y) \ast \text{PSF}(x, y).
\end{equation}

In the encoder stage, we incorporate a Spatial Compression Module (SCM), which leverages Reconstruction Blocks (ReCB) as its primary feature transformation mechanism and employs gating strategies to enhance feature fusion. First, $I_{\text{measurement}}(x, y)$ is processed on multiple scales, extracting spatial context at varying resolutions. These features establish the foundation for reconstructing the spatial representation of the object.

To account for the coded mask distribution associated with the system PSF, we introduce a learnable Coded Mask Simulator (CMS). This module applies a dynamic channel attention mechanism to modulate light intensity across different regions, effectively modeling the coded mask's influence on the system's PSF. By learning the relationship between spatial inputs and the PSF, the CMS dynamically adapts to changing imaging conditions, improving the simulation of light intensity patterns. Its output, a distribution map that encodes high-level semantic cues, facilitates more accurate reconstruction of structures, especially in complex or dynamic scenarios where traditional fixed PSF methods often fail.

We further extract frequency domain features at each scale via the Wiener Fusion Block (WNFB). This block performs a Fast Fourier Transform (FFT) on both the spatial features and the output of the Coded Mask Simulator, and then computes the Wiener filter transfer function:
\begin{equation}
    H(u,v) = \frac{\overline{\operatorname{PSF}(u,v)}}{|\operatorname{PSF}(u,v)|^2 + \delta},
\end{equation}
where $H(u,v)$ denotes the Wiener filter in the frequency domain, $\overline{\operatorname{PSF}(u,v)}$ is the complex conjugate of the PSF, and $|\operatorname{PSF}(u,v)|^2$ is the energy spectrum of the PSF.

Using this transfer function, the encoded frequency-domain signal $B(u,v)$ is preliminarily restored:
\begin{equation}
    I_{\text{restored}}(u,v) = H(u,v) \cdot B(u,v).
\end{equation}

In the decoder stage, we propose the Spatial Amplification Module (SAM) to integrate spatial and frequency domain information more comprehensively. The SAM upsamples low-resolution feature maps from deeper layers via bilinear interpolation and aligns them (through zero-padding) with higher-resolution encoder features. These feature maps are then concatenated to preserve both fine-grained detail and broader contextual cues. Serving as a multi-scale fusion mechanism, the SAM leverages information from both domains to produce high-quality reconstructions, ensuring that critical spatial details and global consistency are well-preserved.

\subsection{Coded Mask Simulator}
As discussed in \cref{sec:pre}, the physical level $\operatorname{PSF}(x, y)$ corresponds to the coded mask placed within the sensor. Traditional methods commonly adopt static or fixed kernels to approximate $\operatorname{PSF}(x, y)$. In contrast, we seek a more flexible approach that adaptively learns these coded patterns, resulting in a learnable PSF and minimizing constraints seen in conventional models.

To address this need, we propose the Coded Mask Simulator (CMS). By employing a channel attention mechanism, the CMS captures the intensity distribution features in $I_{\text{measurement}}(x, y)$, which encode the coded mask pattern. Consequently, the system's $\operatorname{PSF}(x, y)$ can be inferred from these learned mask distributions. This attention mechanism models directional inter-pixel relationships, enabling the network to deconvolve scattering and diffraction artifacts induced by the mask. More concretely, it identifies directional modulations in the image and adaptively recalibrates features according to attention weights, achieving a more robust partial or full recovery of occluded signals.

Formally, let $X \in \mathbb{R}^{N \times C \times H \times W}$ be the input feature map. We first apply global average pooling to aggregate spatial information, preserving only channel-level statistics:
\begin{equation}
    z_c = \frac{1}{H \times W} \sum_{i=1}^{H} \sum_{j=1}^{W} X_{c,i,j}, \quad \forall c \in \{1,2,\dots,C\}.
\end{equation}

Subsequently, a pointwise convolution adjusts the resulting channel weights. This operation, which is more efficient than deeper multilayer perceptrons, preserves channel-specific modulation dynamics:
\begin{equation}
    s_c = \sigma(W \cdot z + b),
\end{equation}
where $W \in \mathbb{R}^{C' \times C'}$ (with $C' = \frac{C}{2}$) is the $1 \times 1$ convolution weight matrix, $\sigma$ is a (linear) activation function, and $s_c \in \mathbb{R}^{N \times C' \times 1 \times 1}$ are the channel attention weights.

In summary, the proposed Coded Mask Simulator models the physical coded mask by learning its intensity distribution and directional modulation patterns from the measurement image. This approach implicitly recovers $\operatorname{PSF}(x, y)$ and provides valuable semantic cues that guide subsequent reconstruction steps. The learned distribution fosters flexible, adaptive, and comprehensive recovery of scene information, overcoming the drawbacks associated with traditional fixed PSF kernels.

\subsection{Objective Function}
We train LensNet by jointly minimizing three complementary loss terms: the mean squared error (MSE), the structural similarity index measure (SSIM) and the perceptual image patch similarity (LPIPS). Formally, we define the overall loss function as:
\begin{equation}
    \mathcal{L} = \mathcal{L}_{\mathrm{MSE}} + 0.2 \cdot \mathcal{L}_{\mathrm{SSIM}} + 0.2 \cdot \mathcal{L}_{\mathrm{LPIPS}},
\end{equation}
where each loss provides a different perspective on reconstruction fidelity.

1. Mean Squared Error (MSE)  
\begin{equation}
    \mathcal{L}_{\mathrm{MSE}} = \frac{1}{N} \sum_{i=1}^{N} \left\lVert \widehat{\mathbf{I}}_i - \mathbf{I}_i \right\rVert^2,
\end{equation}
where $\widehat{\mathbf{I}}_i$ denotes the output image, and $\mathbf{I}_i$ represents the corresponding ground truth. By quantifying pixel-wise intensity differences, MSE imposes strong penalties for large deviations and steadily guides the network toward minimizing low-level reconstruction errors.

2. Structural Similarity Index Measure (SSIM)  
\begin{equation}
    \mathcal{L}_{\mathrm{SSIM}} = 1 - \mathrm{SSIM}(\widehat{\mathbf{I}}_i, \mathbf{I}_i),
\end{equation}

This term evaluates perceptual quality by examining local image structures. SSIM is particularly effective for preserving global consistency and fine structural details, ensuring that the reconstructed images align with human perception of image integrity.

3. Learned Perceptual Image Patch Similarity (LPIPS)  
\begin{equation}
    \mathcal{L}_{\mathrm{LPIPS}} = \mathrm{LPIPS}(\widehat{\mathbf{I}}_i, \mathbf{I}_i).
\end{equation}

LPIPS leverages deep neural network features to evaluate perceptual similarity. Unlike MSE or SSIM, which target lower-level and local structure information, LPIPS operates on higher-level representations, promoting reconstructions that faithfully capture the semantic and perceptual nuances of the scene.

By combining these three loss components, our objective function balances low-level fidelity, structural coherence, and perceptual plausibility. Consequently, this strategy leads to more accurate and visually coherent reconstructions in lensless imaging.

\section{Experiments}
\subsection{Dataset and Preprocessing}
The DiffuserCam dataset~\cite{ref10,ref13} comprises 25,000 pairs of images captured simultaneously by a standard lensed camera (serving as ground truth) and a diffuser-based lensless camera. The raw images, originally at $1080 \times 1920$ pixels, are resized to $480 \times 270$ pixels according to previous methods~\cite{ref8,ref21,ref22,ref29}. Of these 25,000 paired images, 24,000 are allocated for training and 1,000 for testing.

Similarly, the MWDNs dataset~\cite{ref22} consists of 25,000 images displayed on a 27-inch Dell monitor, sourced from the Dogs vs. Cats dataset on Kaggle. Original $1280 \times 1280$ images undergo the official preprocessing steps (zero-padding and resizing) to obtain $320 \times 320$ pixels. This standard preprocessing pipeline ensures consistent input dimensions across both datasets, facilitating robust and fair comparisons of lensless imaging strategies.

\subsection{Implementation Details}
We implemented our method in PyTorch and conducted experiments on 2 NVIDIA RTX 3090 GPUs running Ubuntu. All images were used at their preprocessed resolutions without further downsampling. We set the batch size to 16 for both datasets and applied standard data augmentation strategies, such as random rotation and flipping. For optimization, we employed the Adam optimizer.

\subsection{Evaluation Metrics}
We evaluated the performance of our lensless imaging approach using three widely adopted image quality metrics: PSNR, SSIM, and LPIPS. PSNR quantifies the level of noise reduction, while SSIM measures the preservation of structural and textural fidelity. LPIPS also assesses perceptual similarity by comparing reconstructed images with the ground truth according to human visual perception. Higher PSNR and SSIM values indicate better image quality and structural integrity, while lower LPIPS values suggest improved perceptual fidelity and preservation of fine details.

\subsection{Comparisons with State-of-the-art Methods}
In this section, we compare our proposed LensNet with traditional and learning-based methods to assess reconstruction fidelity, robustness, and overall performance in the DiffuserCam and MWDNs datasets.

\subsubsection{Quantitative Comparisons}
As summarized in \cref{tab:comparison}, traditional algorithms such as Wiener, FISTA, ADMM, and APGD produce relatively low numerical performance when applied to highly multiplexed signals observed in lensless imaging. For instance, in the DiffuserCam dataset, these methods yield PSNR values typically between $7.33$~dB and $13.27$~dB, indicating their limited ability to invert complex degradations. Similarly, iterative optimization algorithms (Vanilla Gradient Descent and Nesterov Gradient Descent) exhibit difficulties in converging to high-quality solutions, particularly in low-light or noisy conditions.

On the other hand, learning-based approaches demonstrate noticeable improvements in reconstruction quality. FlatNet~\cite{ref8} achieves PSNR values of 21.16 dB (DiffuserCam) and 19.08 dB (MWDN). Meanwhile, UDN~\cite{ref21} and TikNet~\cite{ref8} tend to preserve structural signals better than traditional methods, with UDN achieving 26.98 dB in PSNR on MWDNs. MWDN~\cite{ref22}, which uses a multi-scale wavelet decomposition strategy, further improves both PSNR and SSIM.

However, none of the previous methods dominate in every metric. Although UDN excels at preserving edges, it can still miss critical high-frequency texture in challenging illumination patterns. Similarly, though MWDN shows promising fine detail recovery, its susceptibility to high noise levels sometimes results in artifacts or mild oversmoothing under extreme conditions. In contrast, our proposed LensNet consistently outperforms all competing methods across the board, establishing new state-of-the-art results (PSNR: 27.46 dB in DiffuserCam and 33.22 dB in MWDN) alongside substantially lower LPIPS values. And \cref{fig5} shows more intuitively.

\subsubsection{Qualitative Comparisons}
Beyond numerical metrics, we conduct an in-depth qualitative assessment to illustrate how LensNet handles subtle textures, fine details, and complex lighting conditions. \Cref{fig3} presents representative reconstructions from the MWDNs datasets, comparing our method with both traditional and CNN-based baselines, with extended comparisons provided in the Supplementary Material. 

As shown in \cref{fig3}, APGD and other traditional solvers often exhibit visibly blurred regions with pronounced ringing artifacts. FlatNet~\cite{ref8} significantly reduces these artifacts, but introduces oversmoothing, obscuring key high-frequency details such as hair strands or fine-textured fabric. LenslessGAN~\cite{ref29} improves the realism of certain regions; however, it can occasionally generate hallucinated details or struggle with complex boundaries under minimal SNR settings.

By contrast, LensNet consistently recovers sharper edges and more faithful textures, demonstrating well-preserved contours around objects and a closer match to the true intensity profile. As highlighted in the red and green bounding boxes of \cref{fig3}, LensNet excels at reconstructing intricate structures like an animal's fur or subtle shading gradients in natural scenes.


\input{tabletex/2_ab}

\input{figtex/fig5}

\subsection{Ablation Study}
As shown in \cref{tab:ab}, we conducted a detailed ablation study to verify the contributions of different components in our network. A reduced model with only three downsampling layers (ThreeDown) suffers from inadequate reconstruction quality, leading to notable losses in structural and textural details. When the ReconstructionBlock is removed (w/o RecB), the model still fails to preserve crucial local textures necessary for high-fidelity image reconstruction. Even when using the original point spread function (PSF) method, the performance did not exceed that of our learnable PSF design.

Illustrated in \cref{fig4}, the visual results underscore the importance of each architectural component. The ThreeDown variant fails to produce structurally coherent images, while omitting RecB degrades local texture preservation. In contrast, our full model—incorporating both the ReconstructionBlock and the learnable PSF—delivers results that are more faithful to the target images, retaining finer details and exhibiting overall higher visual quality. These observations confirm that each proposed module is crucial for achieving high-accuracy lensless imaging reconstructions.

\section{Conclusion}
In this study, we present LensNet, a novel deep learning framework designed for lensless imaging, which effectively integrates features in both the spatial and frequency domain to address the inherent challenges of lensless image reconstruction. By incorporating a learnable Coded Mask Simulator, LensNet dynamically models optical pathways and the Point Spread Function, enabling the precise and efficient reconstruction of complex scene details, including high-frequency components and fine textures. Our extensive experiments demonstrate that LensNet outperforms state-of-the-art methods, achieving superior image quality across multiple benchmark datasets. Moreover, its robustness in handling noise and variations in the PSF positions LensNet as a promising solution for practical applications in fields such as miniature sensors and medical diagnostics. This work not only enhances the capabilities of lensless imaging systems but also establishes a new standard for image reconstruction in lens-free environments.

\section*{Ethical Statement}
There are no ethical issues.

\section*{Acknowledgments}
This work was supported in part by the Science and Technology Development Fund, Macau SAR, under Grant 0141/2023/RIA2 and 0193/2023/RIA3.

\section*{Contribution Statement}
This work was a collaborative effort by all contributing authors. Xuhang Chen, serving as the corresponding author, is responsible for all communications related to this manuscript.

\bibliographystyle{named}
\bibliography{ijcai25}

\input{supp}

\end{document}

%% file: figtex/fig1.tex
\begin{figure}[ht]
\centering
\includegraphics[width=\columnwidth]{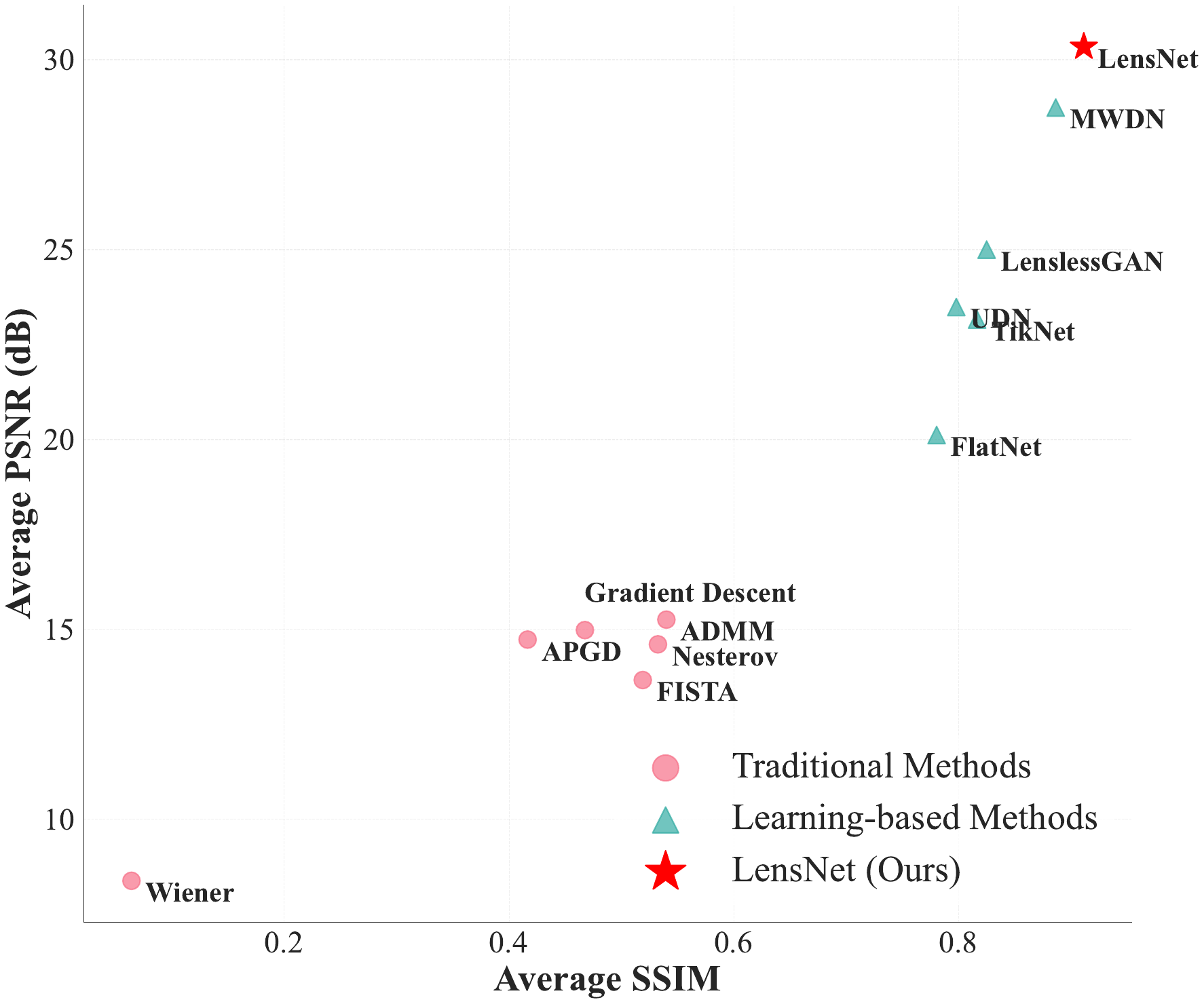} 
\caption{Performance comparison on the MWDNs dataset and the DiffuserCam dataset in terms of PSNR and SSIM.}
\label{fig5}
\end{figure}

%% file: figtex/fig2.tex
\begin{figure*}[ht]
\centering
\includegraphics[width=\textwidth]{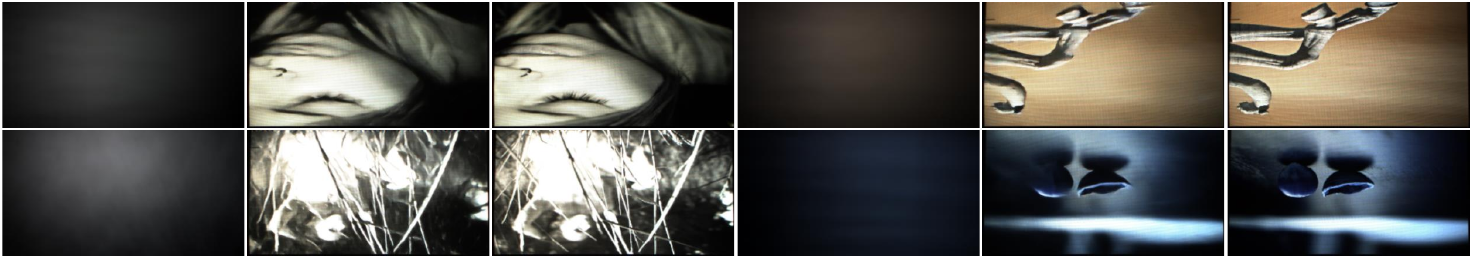} 
\caption{Visualized results on the DiffuserCam dataset. Each image shows the input image, the LensNet reconstruction result, and the corresponding ground truth.}
\label{fig1}
\end{figure*}

%% file: figtex/fig3.tex
\begin{figure*}[ht]
\centering
\includegraphics[width=\textwidth]{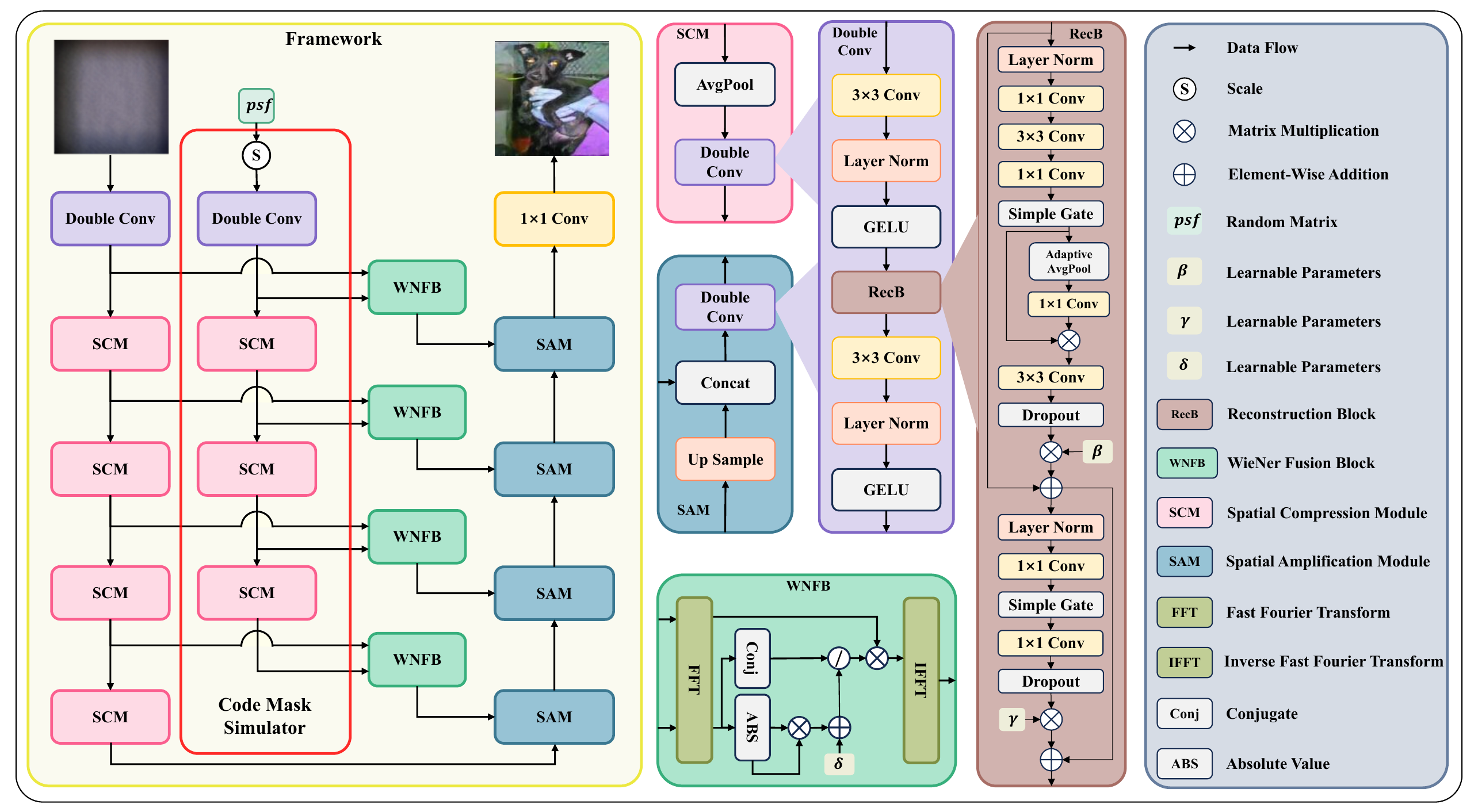}
\caption{The figure illustrates the detailed structure of our model. The input image is measurement on the system and the result is clear scene image through our framework.
}
\label{fig2}
\end{figure*}

%% file: figtex/fig4.tex
\begin{figure*}[ht]
\centering
\includegraphics[width=0.95\textwidth]{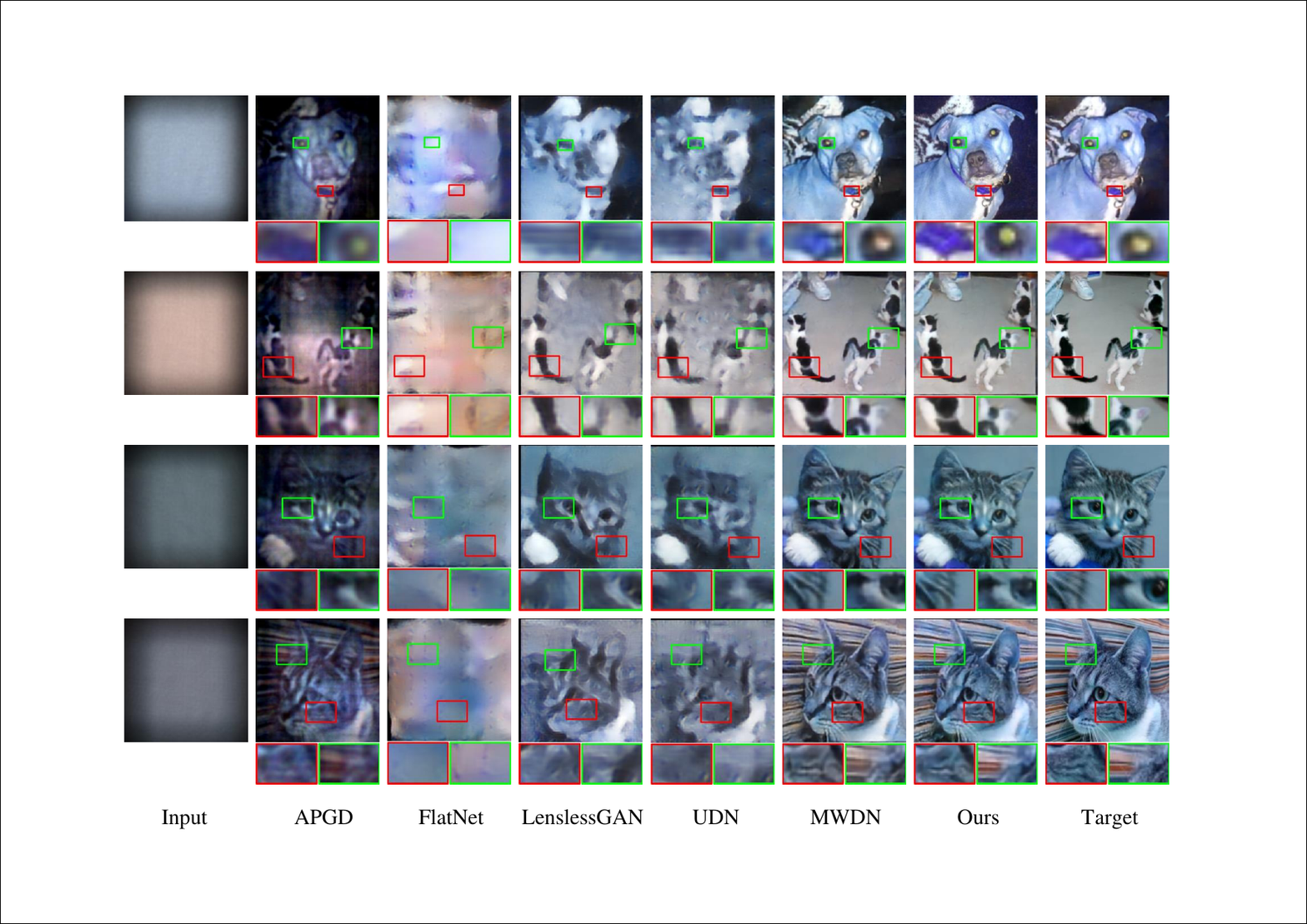} 
\caption{The above shows the qualitative experimental results on MWDNs Dataset. Our method effectively achieves high-quality reconstruction, comparing other methods.}
\label{fig3}
\end{figure*}

%% file: tabletex/1_comp.tex
\begin{table*}[ht]
\centering
\caption{Performance comparison on two datasets: DiffuserCam and MWDNs. Metrics include PSNR (dB), SSIM, and LPIPS. Best results are highlighted in red and bold.}
{
\begin{tabular}{lccc|ccc}
\toprule
\multirow{2}{*}{\textbf{Method}} & \multicolumn{3}{c|}{\textbf{DiffuserCam}} & \multicolumn{3}{c}{\textbf{MWDNs}} \\ \cmidrule(lr){2-7} 
                                 & PSNR$\uparrow$ & SSIM$\uparrow$ & LPIPS$\downarrow$ & PSNR$\uparrow$ & SSIM$\uparrow$ & LPIPS$\downarrow$ \\ \midrule
Wiener~\cite{ref12}
 & 7.33 & 0.083 &  0.770 & 9.44 & 0.045 & 0.731  \\
Vanilla Gradient Descent                         & 13.27 & 0.432 & 0.585  & 16.70 & 0.503 & 0.429 \\
Nesterov Gradient Descent~\cite{ref25} & 12.16 & 0.394 & 0.518 & 17.06 & 0.671  &  0.362\\
FISTA~\cite{ref26} & 11.09 & 0.341 & 0.554 & 16.25 & 0.697 & 0.368 \\
ADMM~\cite{ref28} & 12.76 & 0.442 & 0.541 & 17.76 & 0.638 & 0.343 \\

APGD~\cite{ref27} & 12.13 & 0.385 & 0.518 & 17.34 & 0.448  & 0.439 \\
TikNet~\cite{ref8} & 19.75 & 0.720 & 0.221 & 26.57& 0.913 &0.075 \\
FlatNet~\cite{ref8} & 21.16 & 0.720  & 0.231 & 19.08  & 0.841 & 0.178 \\
LenslessGAN~\cite{ref29} & 22.51 & 0.737 & 0.193 & 27.49 & 0.913 & 0.077\\
UDN~\cite{ref21} & 20.00 & 0.688 & 0.250 & 26.98 & 0.908 & 0.081  \\
MWDN~\cite{ref22}& 25.74 & 0.816 &  0.132 & 31.74 & 0.957& 0.030 \\
\textbf{LensNet}     & \textbf{\textcolor{red}{27.46}}& \textbf{\textcolor{red}{0.863}}& \textbf{\textcolor{red}{0.099}} & \textbf{\textcolor{red}{33.22}} & \textbf{\textcolor{red}{0.960}} & \textbf{\textcolor{red}{0.024}} \\ \bottomrule
\end{tabular}
}
\label{tab:comparison}
\end{table*}

%% file: tabletex/2_ab.tex
\begin{table}[ht]
\centering
\caption{Ablation Study on LensNet}
\small
\begin{tabular}{c|c c c}
\toprule
\multirow{2}{*}{\textbf{Methods}} & \multicolumn{3}{c}{\textbf{LensNet}} \\
 & PSNR$\uparrow$ & SSIM$\uparrow$ & LPIPS$\downarrow$ \\
\midrule
ThreeDown & 25.68 & 0.910 & 0.084 \\
w/o RecB & 30.36 & 0.945  & 0.044\\
w original PSF & 31.92 & 0.954 & 0.031 \\
Ours & \textbf{\textcolor{red}{33.22}} & \textbf{\textcolor{red}{0.960}} & \textbf{\textcolor{red}{0.024}} \\
\bottomrule
\end{tabular}
\label{tab:ab}
\end{table}

%% file: figtex/fig5.tex
\begin{figure}[ht]
\centering
\includegraphics[width=\columnwidth]{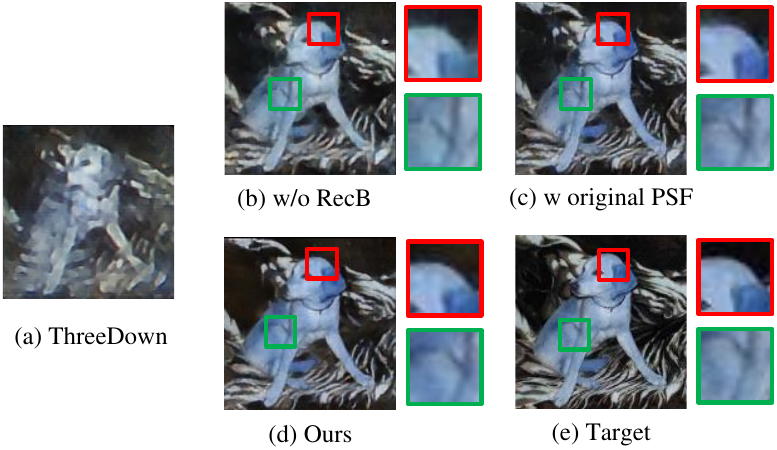} 
\caption{This figure presents the results of our ablation study, highlighting the impact of various components on model performance. 
}
\label{fig4}
\end{figure}

%% file: supp.tex
\appendix
\section{Visualization Result}
\Cref{fig7,fig8,fig9,fig10,fig11} present a comprehensive visual comparison of the most methods discussed in the main paper, providing an extensive performance evaluation. These figures clearly highlight the strengths and limitations of each approach, with particular attention given to the accuracy of reconstruction, the preservation of fine details, and the overall visual quality. Our method demonstrates superior performance across all these dimensions, significantly surpassing both traditional and deep learning-based lensless imaging methods. Notably, it excels not only in overall reconstruction accuracy and detail restoration but also in visual fidelity, ensuring that both structural integrity and fine-grained features are preserved. This advantage is particularly evident in complex scenarios where conventional methods tend to struggle with maintaining image quality in the presence of noise and other artifacts.

\section{User Study}
While objective metrics such as PSNR, SSIM, and LPIPS are commonly used to evaluate lensless imaging methods, they primarily assess pixel-level differences and may not fully capture the human visual perception of image quality. These metrics, though valuable, often overlook subtle nuances such as the preservation of fine details, visual artifacts, and overall perceptual quality. To address this limitation, we designed a user study to complement the objective evaluations, providing a more comprehensive assessment of the visual performance of different imaging methods. This study aims to collect subjective feedback from human observers to gain a deeper understanding of how these methods are perceived in practical, real-world scenarios.

Given that deep learning methods substantially outperform traditional approaches, we limited the participant evaluation to our method and five other learning-based methods to maintain focus and reduce the cognitive load on participants. However, we also included the two most effective traditional methods for comparison. A total of 30 participants were recruited, including both experts in computer vision and non-experts. This diverse sample was chosen to capture a wide range of subjective opinions and preferences, ensuring that the study results reflect general human perception.

A diverse set of lensless imaging datasets was carefully selected, representing a variety of challenging scenarios such as low-light conditions, noise, and distortions. Each image was processed using different lensless imaging methods, resulting in multiple reconstructed versions of each image. We randomly selected 45 images from each
 of the two test sets (DiffuserCam and MWDNs), resulting in a
 total of 90 images for evaluation.

We meticulously prepared the user study form, as presented on the final page of the Word document, ensuring that the images were not compressed. The images were organized into groups, each containing one original input image and eight corresponding lensless imaging results, including our method and seven other traditional and deep learning-based methods. To prevent bias, the methods were anonymized, and the order of the methods within each group was randomized.

Participants were asked to evaluate images reconstructed by different methods. For each image, they provided subjective ratings based on the following refined criteria tailored to lensless imaging:

\begin{itemize}
    \item \textbf{Reconstruction Accuracy}: The degree to which the reconstructed image resembles the original scene, with particular attention to the accuracy of the overall structure and large-scale features.
    
    \item \textbf{Detail Restoration}: The extent to which fine details, such as texture, small edges, and subtle features, are preserved in the image, especially in regions with low contrast or fine-grained structures.
    
    \item \textbf{Artifact Presence}: The visibility and severity of any artifacts introduced during reconstruction, such as noise, blur, ghosting, or ringing, which may distort the image's fidelity.
    
    \item \textbf{Noise Handling}: How well the method mitigates or reduces noise in the reconstructed image, particularly in low-light conditions or areas of high noise.
    
    \item \textbf{Visual Realism}: The naturalness of the image, including the smoothness of the transitions between regions, and the avoidance of unnatural edges or transitions that are common in lensless imaging methods.
    
\end{itemize}

\begin{table*}[ht]
\centering
\begin{tabular}{c|c|c|c|c|c|c|c|c}
\toprule
\textbf{Method} & \textbf{Reconstruction Accuracy} & \textbf{Detail Restoration} & \textbf{Artifact Presence} & \textbf{Noise Handling} & \textbf{Visual Realism} \\ \midrule
\textbf{ADMM}    &   1.88  &  1.70  &  1.60   &  1.59   &   1.68      \\ \midrule
\textbf{APGD}    &   2.52  &  2.34  &   2.09  &  2.08   &    2.42     \\ \midrule
\textbf{TikNet}  &  3.10    &  2.74  &  2.61   &   2.64  &  2.85       \\ \midrule
\textbf{FlatNet} &  1.85   & 1.72   & 1.69    &  1.67   &    1.69     \\ \midrule
\textbf{LenslessGAN} &   2.71 & 2.42 &  2.19 &  2.27   & 2.44  \\ \midrule
\textbf{UDN}     &  2.57 &  2.32  &  2.14  &   2.18 &   2.31        \\ \midrule
\textbf{MWDN}    & 4.12  & 3.96  & 3.88  &  3.86  &   4.01    \\ \midrule
\textbf{Ours}    &   \textcolor{red}{4.20}   &  \textcolor{red}{4.14} &  \textcolor{red}{4.10}   & \textcolor{red}{4.11}   & \textcolor{red}{4.19}   \\ 
\bottomrule
\end{tabular}
\caption{User study results for different methods across multiple evaluation metrics.}
\label{tab:userstudy}
\end{table*}

A five-point Likert scale was used for each criterion (1 being poor, 5 being excellent). In addition to numerical ratings, participants were encouraged to provide qualitative feedback regarding their preferences and any noticeable strengths or weaknesses in the images.

\begin{itemize}
    \item \textbf{1 (Poor)}: The image significantly falls short in the specific criterion, marked by noticeable issues or distortions.
    \item \textbf{2 (Fair)}: The image, despite visible flaws, exhibits some elements of acceptable quality.
    \item \textbf{3 (Average)}: The image is satisfactory overall, with most elements adequately processed.
    \item \textbf{4 (Good)}: The image is well-processed, presenting only minor imperfections.
    \item \textbf{5 (Excellent)}: The image excels in the criterion, demonstrating exceptional quality.
\end{itemize}

The study was performed in a controlled environment to minimize external variables. To ensure fairness, all images were standardized to the same resolution. Participants were instructed to independently rate each image based on the specified criteria, ensuring a comprehensive evaluation of the various aspects of lensless imaging. During the evaluation, participants were allowed to zoom in on the images for a closer inspection. For each image, the original was presented alongside the outputs from the eight methods, with the method names anonymized to prevent bias.

The results of the user study, as summarized in \cref{tab:userstudy}, demonstrate the superiority of our method across all evaluation metrics. Our method consistently outperformed the other techniques in terms of reconstruction accuracy, detail restoration, artifact presence, noise handling, and visual realism. The participants' ratings reflect a clear preference for our approach, indicating its effectiveness in delivering high-quality lensless imaging results. These findings highlight the robustness and practicality of our method in addressing the challenges inherent in lensless imaging, positioning it as a promising solution for future applications in this domain.

\section{Model Analysis}
The main text highlights the strong reconstruction performance of our model. To provide a more comprehensive evaluation, we analyze additional aspects of the model’s performance, including its size, training stability, and robustness.

\subsection{Model Complexity}
We analyze its complexity in terms of parameter count and computational cost (FLOPs). The tensors we tested are the same as the image sizes in the MWDNs dataset. As shown in \cref{tab:param_flops}, UDN demonstrates the most lightweight design, with only 1.04M parameters and 2.37G FLOPs. In contrast, FlatNet and TikNet exhibit the highest complexity, requiring approximately 59M parameters and 220G FLOPs. Our method achieves a balanced trade-off, with 31.18M parameters and 115.10G FLOPs, offering a compromise between model capacity and computational efficiency. LenslessGAN also exhibits good efficiency, featuring a relatively low parameter count (11.77M) and computational cost (13.82G FLOPs). These results provide valuable insight into the practical implementation considerations of various methods.

\begin{table}[h]
\centering
\caption{Comparison of Parameter Count and FLOPs}
\begin{tabular}{lcc}
\toprule
\textbf{Method} & \textbf{Params (M)} & \textbf{FLOPs (G)} \\ \midrule
Ours            & 31.18                 & 115.10               \\
FlatNet          & 59.13                 & 220.42               \\
LenslessGAN       & 11.77                 & 13.82               \\
MWDN          & 21.96                 & 83.85               \\ 
TikNet          & 59.13                 & 220.42               \\ 
UDN          & 1.04                 & 2.37               \\ 
\bottomrule
\end{tabular}
\label{tab:param_flops}
\end{table}

\subsection{Training Stability}
One significant advantage of our model is its remarkable training stability compared to other deep learning models. Across multiple runs, our model consistently converges without issues such as loss divergence or NaN values. In contrast, many existing models in this domain often experience instability during training, where the loss either fails to converge or becomes undefined. This stability stems from the careful architectural design and optimization strategies integrated into our approach. In particular, our model addresses the unique challenges of lensless imaging, including the inherent ill-posed nature of the inverse problem and the sensitivity to noise and perturbations in the data. By incorporating features that enhance gradient flow stability and numerical precision, our method reliably converges even under these challenging conditions. Such stability is critical for practical applications, where consistent training outcomes are essential for real-world deployment.

\subsection{Robustness}
\begin{figure}[ht]
\centering
\includegraphics[width=\columnwidth]{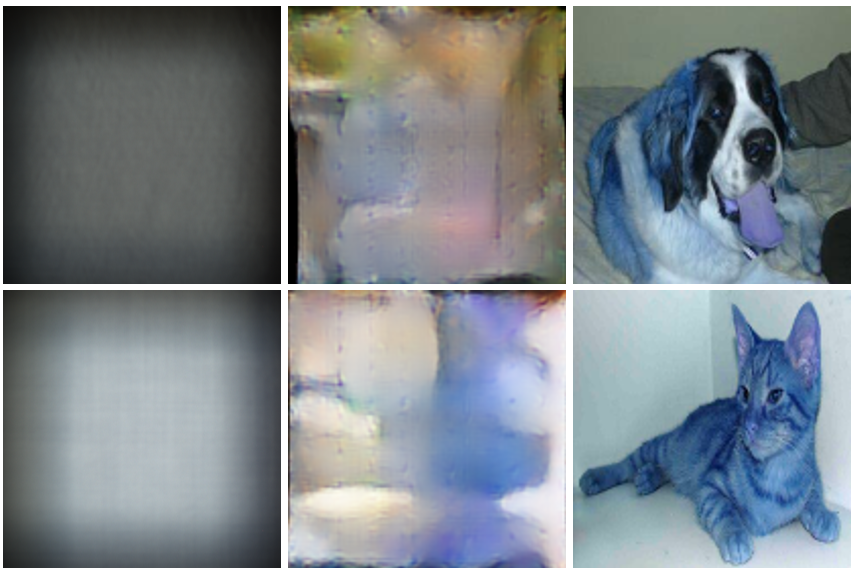} 
\caption{Left image is input,and right image is target. The central image is the result of the FlatNet without PSF.}
\label{fig6}
\end{figure}

\begin{figure*}[ht]
\centering
\includegraphics[width=\textwidth, height=\textheight]{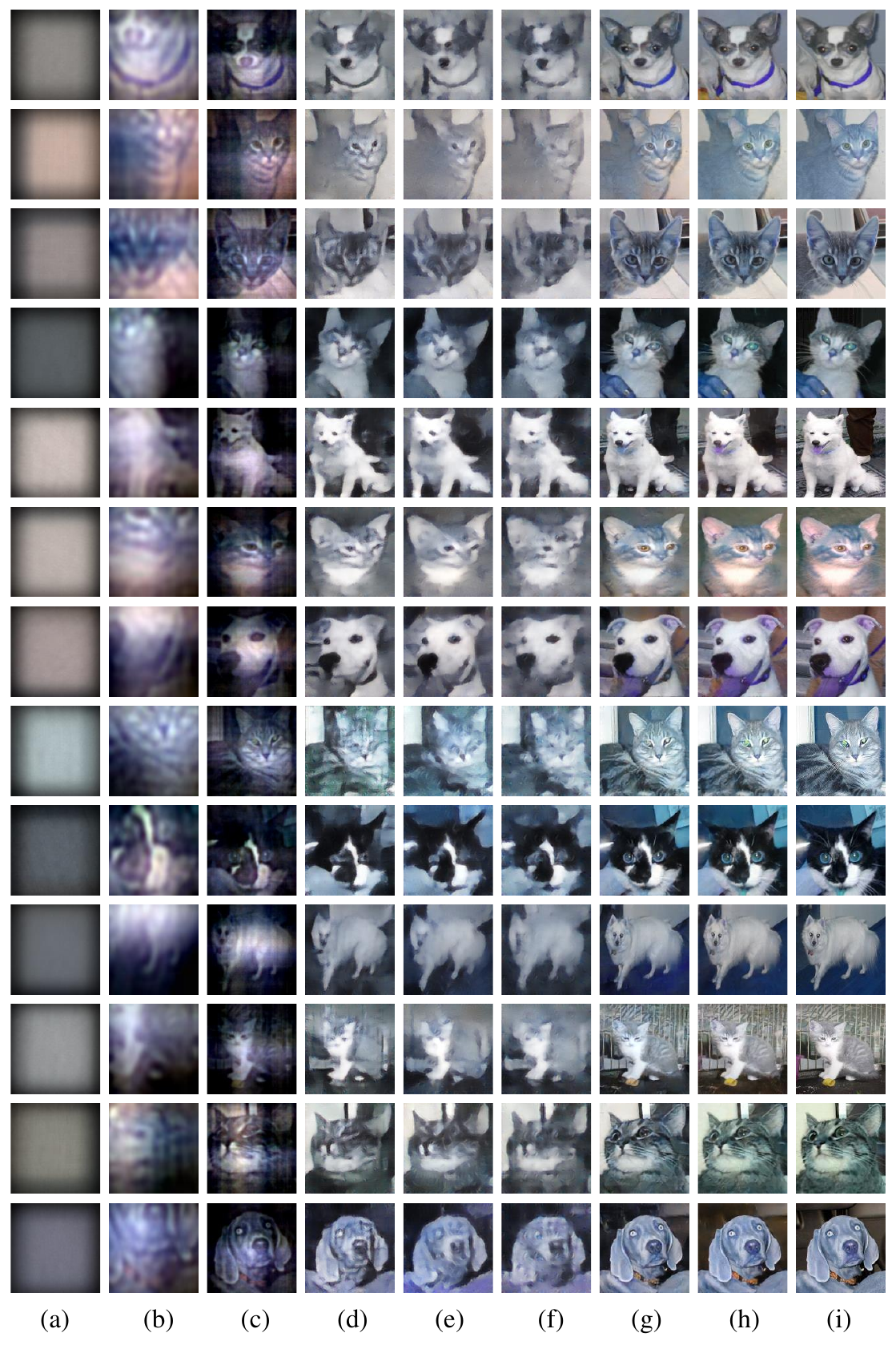} 
\caption{Comprehensive visual comparison on MWDNs Dataset. (a) Input (b) ADMM , (c) APGD , (d) TikNet, (e) LenslessGAN,(f) UDN, (g) MWDN, (h) \textcolor{red}{Ours} , (i) Target}
\label{fig7}
\end{figure*}

\begin{figure*}[ht]
\centering
\includegraphics[width=\textwidth]{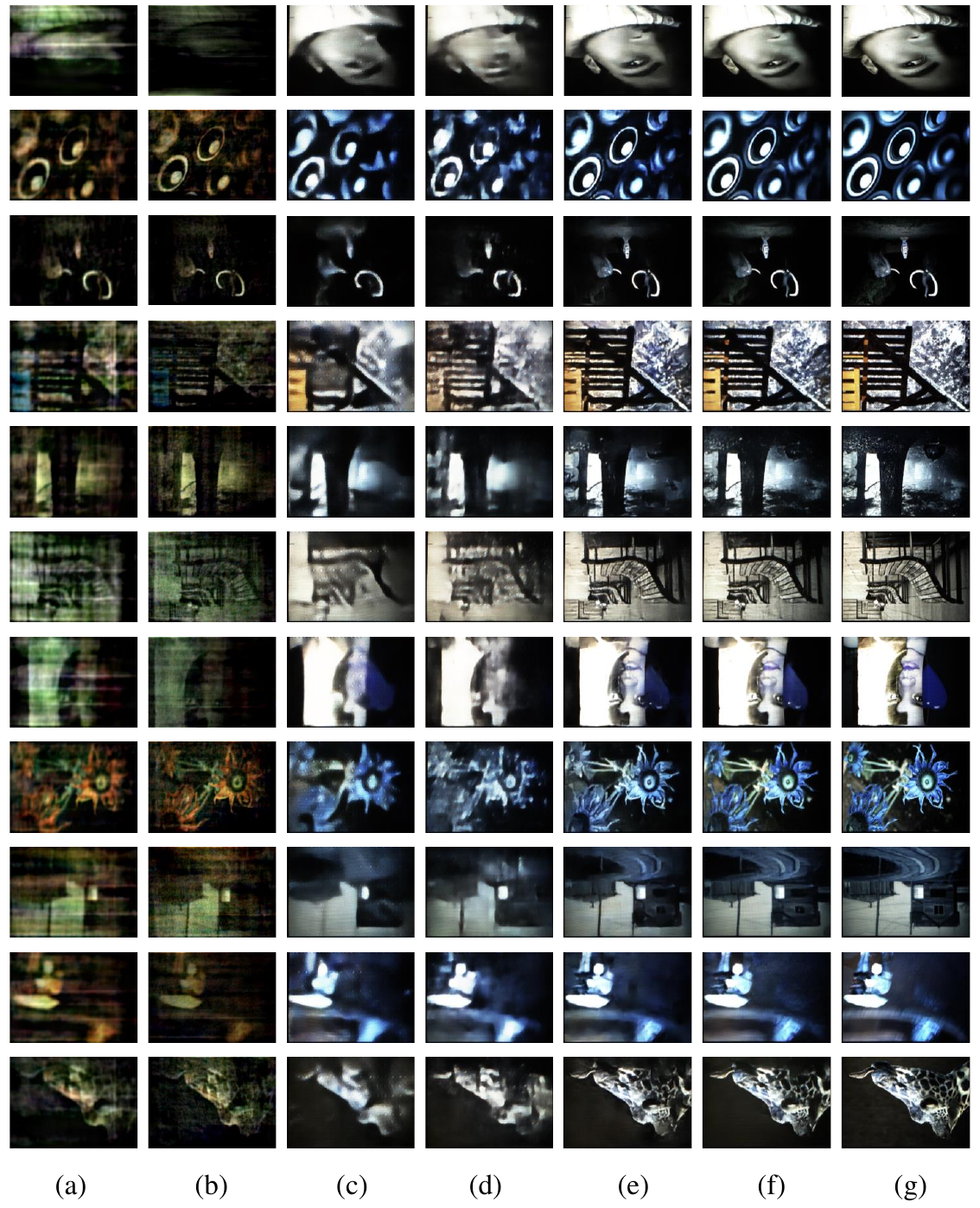} 
\caption{Comprehensive visual comparison on DiffuserCam Dataset. (a) ADMM , (b) APGD , (c) FlatNet, (d) UDN, (e) MWDN, (f) \textcolor{red}{Ours} , (g) Target}
\label{fig8}
\end{figure*}

\begin{figure*}[ht]
\centering
\includegraphics[width=\textwidth, height=\textheight]{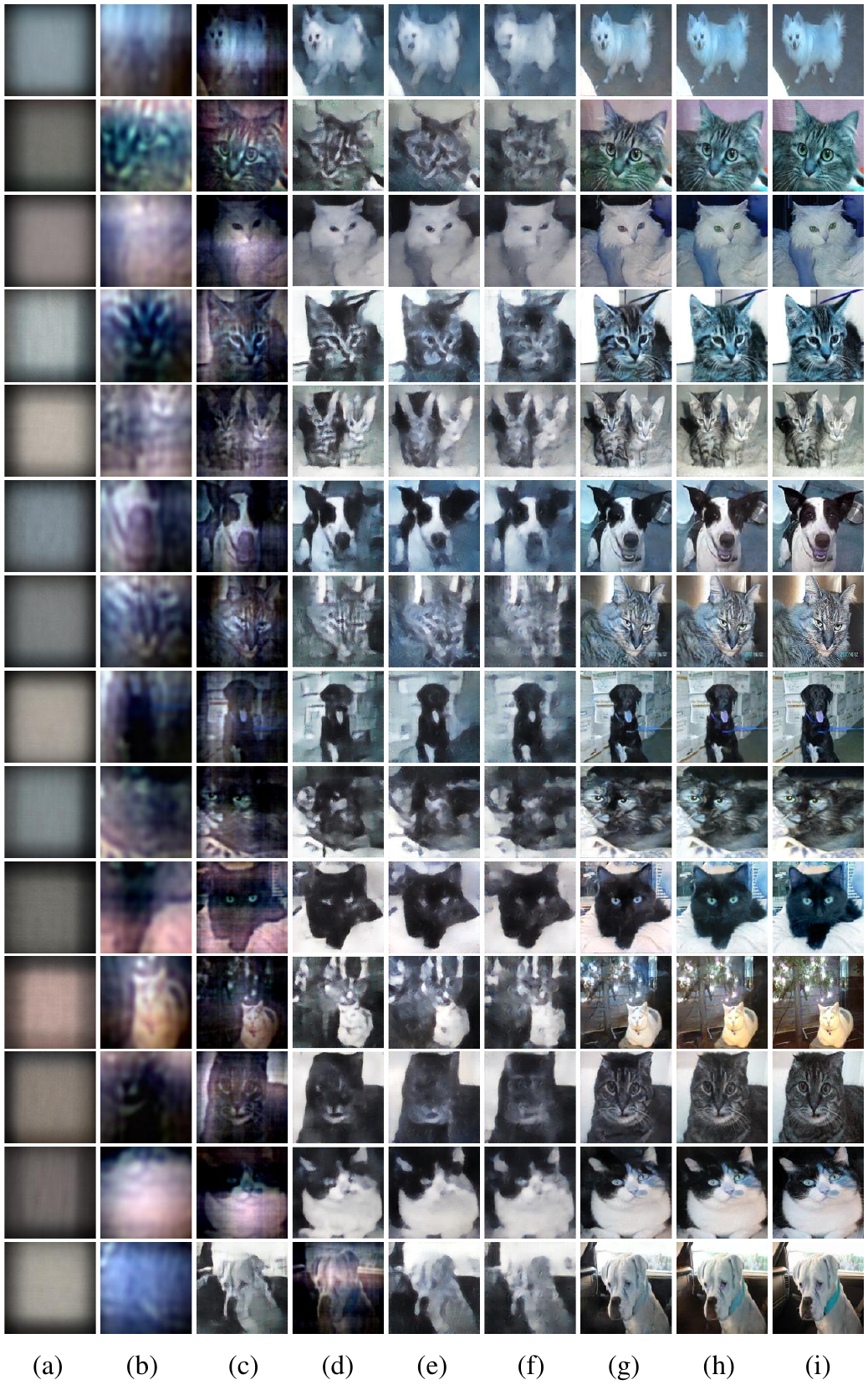} 
\caption{Comprehensive visual comparison on MWDNs Dataset. (a) Input (b) ADMM , (c) APGD , (d) TikNet, (e) LenslessGAN,(f) UDN, (g) MWDN, (h) \textcolor{red}{Ours} , (i) Target}
\label{fig9}
\end{figure*}

\begin{figure*}[ht]
\centering
\includegraphics[width=\textwidth, height=\textheight]{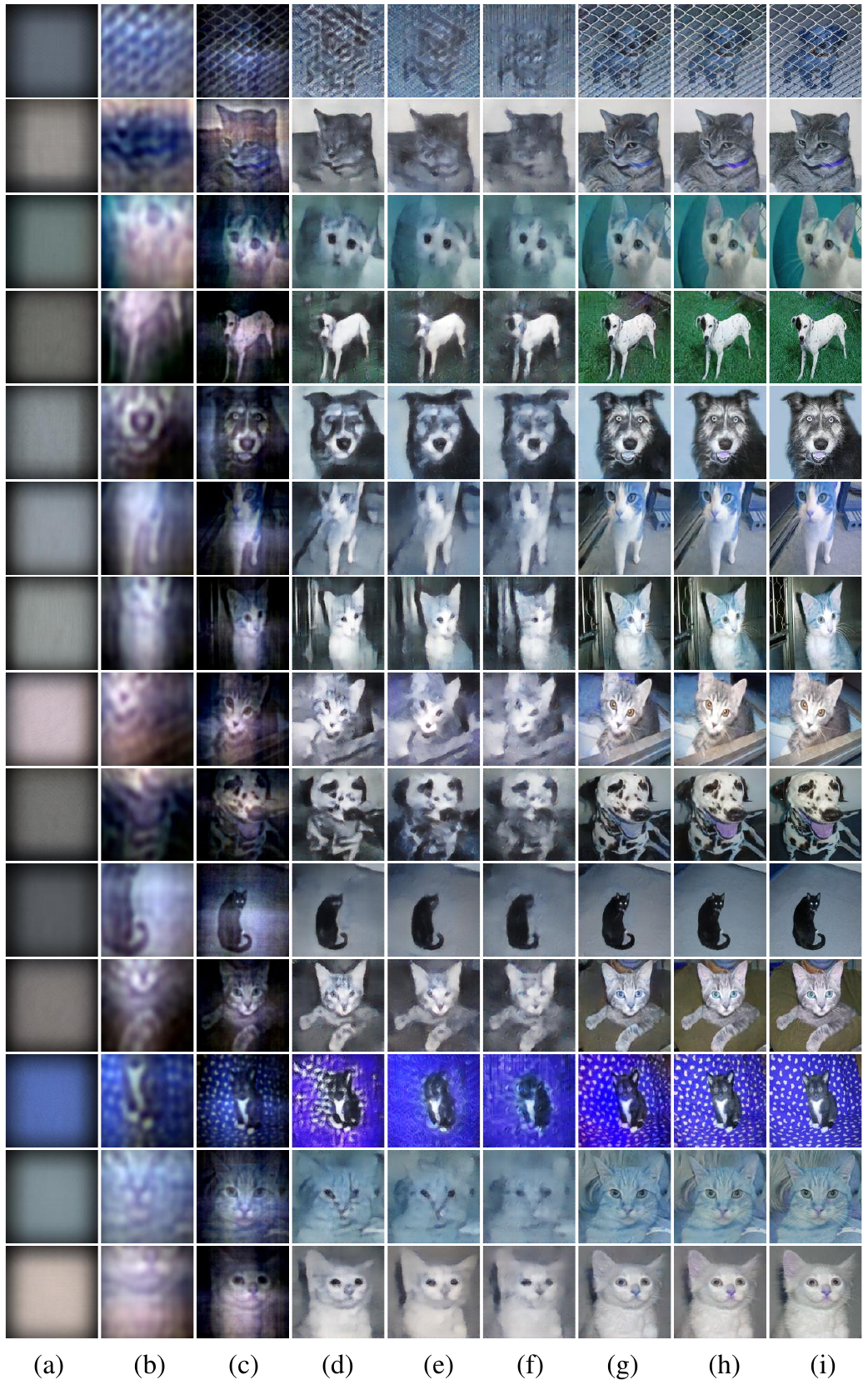} 
\caption{Comprehensive visual comparison on MWDNs Dataset. (a) Input (b) ADMM , (c) APGD , (d) TikNet, (e) LenslessGAN,(f) UDN, (g) MWDN, (h) \textcolor{red}{Ours} , (i) Target}
\label{fig10}
\end{figure*}

\begin{figure*}[ht]
\centering
\includegraphics[width=\textwidth]{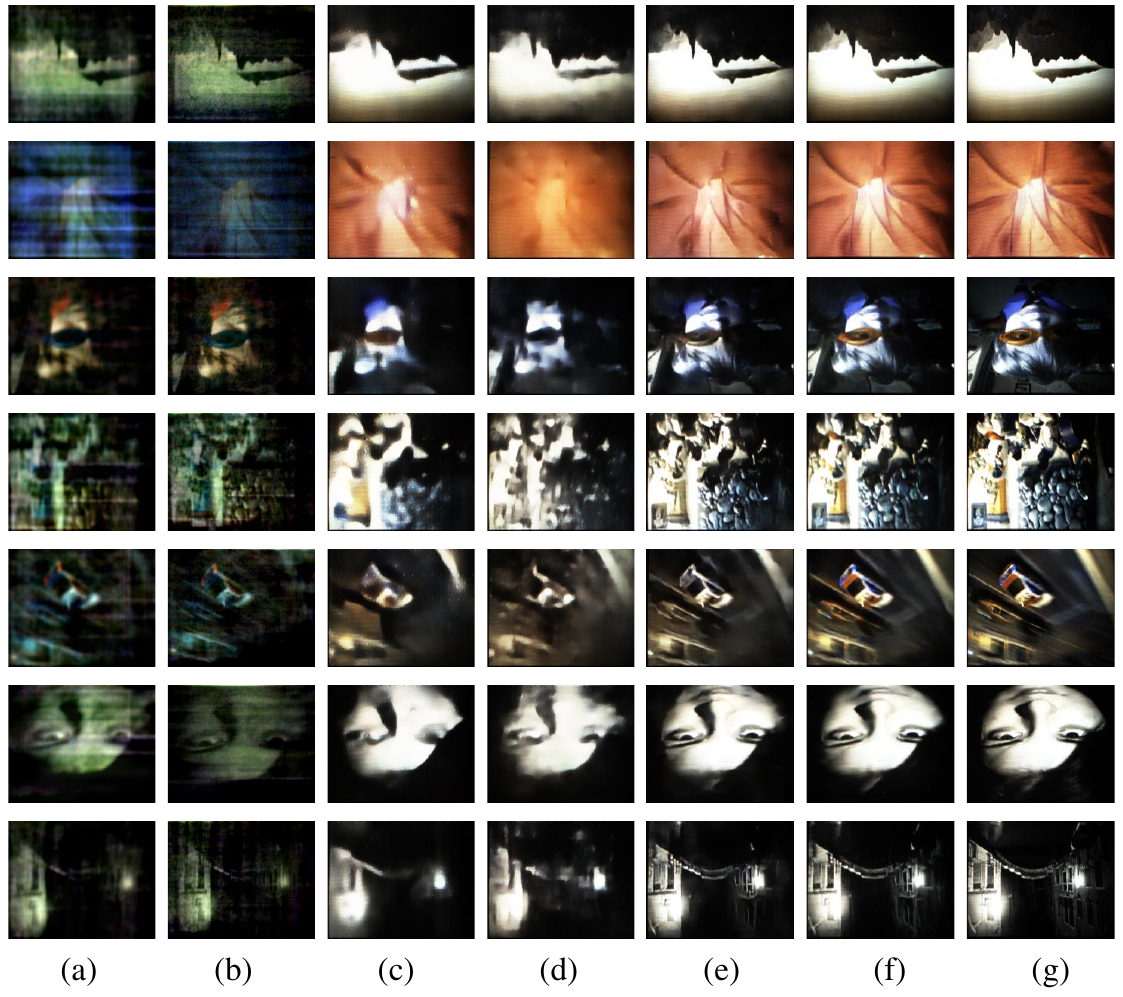} 
\caption{Comprehensive visual comparison on DiffuserCam Dataset. (a) ADMM , (b) APGD , (c) FlatNet, (d) UDN, (e) MWDN, (f) \textcolor{red}{Ours} , (g) Target}
\label{fig11}
\end{figure*}

Our model demonstrates enhanced robustness compared to traditional deep learning approaches in lensless imaging. Conventional methods rely heavily on prior knowledge of the point spread function (PSF) for image reconstruction, such as \cref{fig6} , making their performance highly sensitive to errors in PSF estimation. In contrast, our model learns an adaptive PSF directly from the data, eliminating the need for predefined PSF knowledge. This data-driven approach inherently adapts to variations in optical setups and environmental conditions, allowing the model to generalize effectively across a wide range of scenarios. Additionally, by bypassing explicit PSF modeling, our method mitigates error propagation caused by inaccurate PSF estimates, thereby improving the overall reliability of the reconstruction process. Such adaptability is particularly beneficial for practical applications, where PSF characteristics may fluctuate due to manufacturing variances or environmental changes.

%% file: ijcai25.bbl
\begin{thebibliography}{}

\bibitem[\protect\citeauthoryear{Antipa \bgroup \em et al.\egroup }{2017}]{ref10}
Nick Antipa, Grace Kuo, Reinhard Heckel, Ben Mildenhall, Emrah Bostan, Ren Ng, and Laura Waller.
\newblock Diffusercam: lensless single-exposure 3d imaging.
\newblock {\em Optica}, 5(1):1--9, 2017.

\bibitem[\protect\citeauthoryear{Asif \bgroup \em et al.\egroup }{2015}]{ref14}
M~Salman Asif, Ali Ayremlou, Ashok Veeraraghavan, Richard Baraniuk, and Aswin Sankaranarayanan.
\newblock Flatcam: Replacing lenses with masks and computation.
\newblock In {\em ICCV Workshop}, pages 663--666. IEEE, 2015.

\bibitem[\protect\citeauthoryear{Asif \bgroup \em et al.\egroup }{2016}]{ref9}
M~Salman Asif, Ali Ayremlou, Aswin Sankaranarayanan, Ashok Veeraraghavan, and Richard~G Baraniuk.
\newblock Flatcam: Thin, lensless cameras using coded aperture and computation.
\newblock {\em TCI}, 3(3):384--397, 2016.

\bibitem[\protect\citeauthoryear{Beck and Teboulle}{2009a}]{ref2}
Amir Beck and Marc Teboulle.
\newblock A fast iterative shrinkage-thresholding algorithm for linear inverse problems.
\newblock {\em SIAM journal on imaging sciences}, 2(1):183--202, 2009.

\bibitem[\protect\citeauthoryear{Beck and Teboulle}{2009b}]{ref26}
Amir Beck and Marc Teboulle.
\newblock A fast iterative shrinkage-thresholding algorithm for linear inverse problems.
\newblock {\em SIAM journal on imaging sciences}, 2(1):183--202, 2009.

\bibitem[\protect\citeauthoryear{Boominathan \bgroup \em et al.\egroup }{2020}]{ref11}
Vivek Boominathan, Jesse~K Adams, Jacob~T Robinson, and Ashok Veeraraghavan.
\newblock Phlatcam: Designed phase-mask based thin lensless camera.
\newblock {\em TPAMI}, 42(7):1618--1629, 2020.

\bibitem[\protect\citeauthoryear{Boyd \bgroup \em et al.\egroup }{2011}]{ref28}
Stephen Boyd, Neal Parikh, Eric Chu, Borja Peleato, Jonathan Eckstein, et~al.
\newblock Distributed optimization and statistical learning via the alternating direction method of multipliers.
\newblock {\em Foundations and Trends{\textregistered} in Machine learning}, 3(1):1--122, 2011.

\bibitem[\protect\citeauthoryear{Cai \bgroup \em et al.\egroup }{2024}]{ref7}
Xin Cai, Zhiyuan You, Hailong Zhang, Wentao Liu, Jinwei Gu, and Tianfan Xue.
\newblock Phocolens: Photorealistic and consistent reconstruction in lensless imaging.
\newblock {\em arXiv}, 2024.

\bibitem[\protect\citeauthoryear{Chen \bgroup \em et al.\egroup }{2023}]{chen8}
Xuhang Chen, Baiying Lei, Chi-Man Pun, and Shuqiang Wang.
\newblock Brain diffuser: An end-to-end brain image to brain network pipeline.
\newblock In {\em PRCV}, pages 16--26, 2023.

\bibitem[\protect\citeauthoryear{Chen \bgroup \em et al.\egroup }{2024}]{chen11}
Xuhang Chen, Chi-Man Pun, and Shuqiang Wang.
\newblock Medprompt: Cross-modal prompting for multi-task medical image translation.
\newblock In {\em PRCV}, pages 61--75, 2024.

\bibitem[\protect\citeauthoryear{Chen \bgroup \em et al.\egroup }{2025}]{chen7}
Guanli Chen, Guoheng Huang, Xiaochen Yuan, Xuhang Chen, Guo Zhong, and Chi-Man Pun.
\newblock Cross-view geo-localization via learning correspondence semantic similarity knowledge.
\newblock In {\em MMM}, pages 220--233, 2025.

\bibitem[\protect\citeauthoryear{Goodman}{2005}]{ref23}
Joseph~W Goodman.
\newblock {\em Introduction to Fourier optics}.
\newblock Roberts and Company publishers, 2005.

\bibitem[\protect\citeauthoryear{Guo \bgroup \em et al.\egroup }{2024}]{chen4}
Xiaojiao Guo, Xuhang Chen, Shenghong Luo, Shuqiang Wang, and Chi-Man Pun.
\newblock Dual-hybrid attention network for specular highlight removal.
\newblock In {\em ACM MM}, pages 10173--10181, 2024.

\bibitem[\protect\citeauthoryear{Guo \bgroup \em et al.\egroup }{2025a}]{chen2}
Xiaojiao Guo, Xuhang Chen, Shuqiang Wang, and Chi-Man Pun.
\newblock Underwater image restoration through a prior guided hybrid sense approach and extensive benchmark analysis.
\newblock {\em TCSVT}, 2025.

\bibitem[\protect\citeauthoryear{Guo \bgroup \em et al.\egroup }{2025b}]{chen5}
Xiaojiao Guo, Yihang Dong, Xuhang Chen, Weiwen Chen, Zimeng Li, FuChen Zheng, and Chi-Man Pun.
\newblock Underwater image restoration via polymorphic large kernel cnns.
\newblock In {\em ICASSP}, pages 1--5, 2025.

\bibitem[\protect\citeauthoryear{Huo \bgroup \em et al.\egroup }{2024}]{chen13}
Yejing Huo, Guoheng Huang, Lianglun Cheng, Jianbin He, Xuhang Chen, Xiaochen Yuan, Guo Zhong, and Chi-Man Pun.
\newblock Iman: An adaptive network for robust npc mortality prediction with missing modalities.
\newblock In {\em BIBM}, pages 2074--2079, 2024.

\bibitem[\protect\citeauthoryear{Khan \bgroup \em et al.\egroup }{2019}]{ref16}
Salman~S Khan, VR~Adarsh, Vivek Boominathan, Jasper Tan, Ashok Veeraraghavan, and Kaushik Mitra.
\newblock Towards photorealistic reconstruction of highly multiplexed lensless images.
\newblock In {\em ICCV}, pages 7860--7869, 2019.

\bibitem[\protect\citeauthoryear{Khan \bgroup \em et al.\egroup }{2020}]{ref8}
Salman~Siddique Khan, Varun Sundar, Vivek Boominathan, Ashok Veeraraghavan, and Kaushik Mitra.
\newblock Flatnet: Towards photorealistic scene reconstruction from lensless measurements.
\newblock {\em TPAMI}, 44(4):1934--1948, 2020.

\bibitem[\protect\citeauthoryear{Kingshott \bgroup \em et al.\egroup }{2022}]{ref21}
Oliver Kingshott, Nick Antipa, Emrah Bostan, and Kaan Ak{\c{s}}it.
\newblock Unrolled primal-dual networks for lensless cameras.
\newblock {\em Optics Express}, 30(26):46324--46335, 2022.

\bibitem[\protect\citeauthoryear{Li and Lin}{2015}]{ref27}
Huan Li and Zhouchen Lin.
\newblock Accelerated proximal gradient methods for nonconvex programming.
\newblock {\em NeurIPS}, 28, 2015.

\bibitem[\protect\citeauthoryear{Li \bgroup \em et al.\egroup }{2023a}]{ref22}
Ying Li, Zhengdai Li, Kaiyu Chen, Youming Guo, and Changhui Rao.
\newblock Mwdns: reconstruction in multi-scale feature spaces for lensless imaging.
\newblock {\em Optics Express}, 31(23):39088--39101, 2023.

\bibitem[\protect\citeauthoryear{Li \bgroup \em et al.\egroup }{2023b}]{chen3}
Zinuo Li, Xuhang Chen, Chi-Man Pun, and Xiaodong Cun.
\newblock High-resolution document shadow removal via a large-scale real-world dataset and a frequency-aware shadow erasing net.
\newblock In {\em ICCV}, pages 12449--12458, 2023.

\bibitem[\protect\citeauthoryear{Li \bgroup \em et al.\egroup }{2023c}]{chen10}
Zinuo Li, Xuhang Chen, Shuqiang Wang, and Chi-Man Pun.
\newblock A large-scale film style dataset for learning multi-frequency driven film enhancement.
\newblock In {\em IJCAI}, pages 1160--1168, 2023.

\bibitem[\protect\citeauthoryear{Li \bgroup \em et al.\egroup }{2024a}]{chen1}
Mingxian Li, Hao Sun, Yingtie Lei, Xiaofeng Zhang, Yihang Dong, Yilin Zhou, Zimeng Li, and Xuhang Chen.
\newblock High-fidelity document stain removal via a large-scale real-world dataset and a memory-augmented transformer.
\newblock In {\em WACV}, pages 7614--7624, 2024.

\bibitem[\protect\citeauthoryear{Li \bgroup \em et al.\egroup }{2024b}]{li2024cross}
Xiaohong Li, Guoheng Huang, Lianglun Cheng, Guo Zhong, Weihuang Liu, Xuhang Chen, and Muyan Cai.
\newblock Cross-domain visual prompting with spatial proximity knowledge distillation for histological image classification.
\newblock {\em Journal of Biomedical Informatics}, 158:104728, 2024.

\bibitem[\protect\citeauthoryear{Liu \bgroup \em et al.\egroup }{2023a}]{ref18}
Muyuan Liu, Xiuqin Su, Xiaopeng Yao, Wei Hao, and Wenhua Zhu.
\newblock Lensless image restoration based on multi-stage deep neural networks and pix2pix architecture.
\newblock In {\em Photonics}, volume~10, page 1274. MDPI, 2023.

\bibitem[\protect\citeauthoryear{Liu \bgroup \em et al.\egroup }{2023b}]{liu2023coordfill}
Weihuang Liu, Xiaodong Cun, Chi-Man Pun, Menghan Xia, Yong Zhang, and Jue Wang.
\newblock Coordfill: Efficient high-resolution image inpainting via parameterized coordinate querying.
\newblock In {\em AAAI}, volume~37, pages 1746--1754, 2023.

\bibitem[\protect\citeauthoryear{Liu \bgroup \em et al.\egroup }{2023c}]{liu2023explicit}
Weihuang Liu, Xi~Shen, Chi-Man Pun, and Xiaodong Cun.
\newblock Explicit visual prompting for low-level structure segmentations.
\newblock In {\em CVPR}, pages 19434--19445, 2023.

\bibitem[\protect\citeauthoryear{Liu \bgroup \em et al.\egroup }{2024}]{liu2024depth}
Weihuang Liu, Xi~Shen, Haolun Li, Xiuli Bi, Bo~Liu, Chi-Man Pun, and Xiaodong Cun.
\newblock Depth-aware test-time training for zero-shot video object segmentation.
\newblock In {\em CVPR}, pages 19218--19227, 2024.

\bibitem[\protect\citeauthoryear{Luo \bgroup \em et al.\egroup }{2024}]{chen9}
Shenghong Luo, Xuhang Chen, Weiwen Chen, Zinuo Li, Shuqiang Wang, and Chi-Man Pun.
\newblock Devignet: High-resolution vignetting removal via a dual aggregated fusion transformer with adaptive channel expansion.
\newblock In {\em AAAI}, pages 4000--4008, 2024.

\bibitem[\protect\citeauthoryear{Monakhova \bgroup \em et al.\egroup }{2019}]{ref13}
Kristina Monakhova, Joshua Yurtsever, Grace Kuo, Nick Antipa, Kyrollos Yanny, and Laura Waller.
\newblock Learned reconstructions for practical mask-based lensless imaging.
\newblock {\em Optics express}, 27(20):28075--28090, 2019.

\bibitem[\protect\citeauthoryear{Nesterov}{1998}]{ref25}
Yurii Nesterov.
\newblock Introductory lectures on convex programming volume i: Basic course.
\newblock {\em Lecture notes}, 3(4):5, 1998.

\bibitem[\protect\citeauthoryear{Pan \bgroup \em et al.\egroup }{2024}]{yao2}
Xiaochao Pan, Jiawei Yao, Hongrui Kou, Tong Wu, and Canran Xiao.
\newblock Harmonicnerf: Geometry-informed synthetic view augmentation for 3d scene reconstruction in driving scenarios.
\newblock In {\em Proceedings of the 32nd ACM International Conference on Multimedia}, pages 5987--5996, 2024.

\bibitem[\protect\citeauthoryear{Rego \bgroup \em et al.\egroup }{2021}]{ref29}
Joshua~D Rego, Karthik Kulkarni, and Suren Jayasuriya.
\newblock Robust lensless image reconstruction via psf estimation.
\newblock In {\em WACV}, pages 403--412, 2021.

\bibitem[\protect\citeauthoryear{Tanida \bgroup \em et al.\egroup }{2001}]{ref15}
Jun Tanida, Tomoya Kumagai, Kenji Yamada, Shigehiro Miyatake, Kouichi Ishida, Takashi Morimoto, Noriyuki Kondou, Daisuke Miyazaki, and Yoshiki Ichioka.
\newblock Thin observation module by bound optics (tombo): concept and experimental verification.
\newblock {\em Applied optics}, 40(11):1806--1813, 2001.

\bibitem[\protect\citeauthoryear{Wahlberg \bgroup \em et al.\egroup }{2012}]{ref5}
Bo~Wahlberg, Stephen Boyd, Mariette Annergren, and Yang Wang.
\newblock An admm algorithm for a class of total variation regularized estimation problems.
\newblock {\em IFAC Proceedings Volumes}, 45(16):83--88, 2012.

\bibitem[\protect\citeauthoryear{Wang \bgroup \em et al.\egroup }{2019}]{ref1}
Yu~Wang, Wotao Yin, and Jinshan Zeng.
\newblock Global convergence of admm in nonconvex nonsmooth optimization.
\newblock {\em Journal of Scientific Computing}, 78:29--63, 2019.

\bibitem[\protect\citeauthoryear{Wiener}{1949}]{ref12}
Norbert Wiener.
\newblock {\em Extrapolation, interpolation, and smoothing of stationary time series: with engineering applications}.
\newblock The MIT press, 1949.

\bibitem[\protect\citeauthoryear{Yang \bgroup \em et al.\egroup }{2025}]{chen12}
Jietao Yang, Guoheng Huang, Yanzhang Luo, Xiaofeng Zhang, Xiaochen Yuan, Xuhang Chen, Chi-Man Pun, and Mu-yan Cai.
\newblock Tcia: A transformer-cnn model with illumination adaptation for enhancing cell image saliency and contrast.
\newblock {\em TIM}, 2025.

\bibitem[\protect\citeauthoryear{Yao \bgroup \em et al.\egroup }{2024}]{yao1}
Jiawei Yao, Chuming Li, and Canran Xiao.
\newblock Swift sampler: Efficient learning of sampler by 10 parameters.
\newblock {\em arXiv preprint arXiv:2410.05578}, 2024.

\bibitem[\protect\citeauthoryear{Zeng and Lam}{2021}]{ref17}
Tianjiao Zeng and Edmund~Y Lam.
\newblock Robust reconstruction with deep learning to handle model mismatch in lensless imaging.
\newblock {\em TCI}, 7:1080--1092, 2021.

\bibitem[\protect\citeauthoryear{Zheng \bgroup \em et al.\egroup }{2024}]{zheng2024smaformer}
Fuchen Zheng, Xuhang Chen, Weihuang Liu, Haolun Li, Yingtie Lei, Jiahui He, Chi-Man Pun, and Shounjun Zhou.
\newblock Smaformer: Synergistic multi-attention transformer for medical image segmentation.
\newblock In {\em BIBM}, 2024.

\bibitem[\protect\citeauthoryear{Zhou \bgroup \em et al.\egroup }{2024}]{chen6}
Ziyang Zhou, Yingtie Lei, Xuhang Chen, Shenghong Luo, Wenjun Zhang, Chi-Man Pun, and Zhen Wang.
\newblock Docdeshadower: Frequency-aware transformer for document shadow removal.
\newblock In {\em SMC}, pages 2468--2473, 2024.

\bibitem[\protect\citeauthoryear{Zhu \bgroup \em et al.\egroup }{2024a}]{chen14}
Dingzhou Zhu, Guoheng Huang, Xiaochen Yuan, Xuhang Chen, Guo Zhong, Chi-Man Pun, and Jie Deng.
\newblock Faqnet: Frequency-aware quaternion network for endoscopic highlight removal.
\newblock In {\em BIBM}, pages 1408--1413, 2024.

\bibitem[\protect\citeauthoryear{Zhu \bgroup \em et al.\egroup }{2024b}]{zhu2024test}
Leyi Zhu, Weihuang Liu, Xinyi Chen, Zimeng Li, Xuhang Chen, Zhen Wang, and Chi-Man Pun.
\newblock Test-time intensity consistency adaptation for shadow detection.
\newblock {\em arXiv}, 2024.

\end{thebibliography}
